\documentclass[aps,prl,twocolumn,superscriptaddress,longbibliography,footinbib]{revtex4-1}  
\usepackage{graphicx} 
\usepackage{bm}
\usepackage{amsmath,amssymb,dsfont,amstext,amsfonts}
\usepackage[colorlinks=true,linkcolor=blue,urlcolor=blue,citecolor=blue]{hyperref}

\begin{document}
\title{$\pi$-fluxes, semi-metals and flat bands in artificial materials}
 \author{Toshikaze Kariyado} 
 \affiliation{International Center for Materials Nanoarchitectonics(WPI-MANA),National Institute for Materials Science, Tsukuba, Ibaraki 305-0044, Japan}
 \affiliation{Department of Physics, Harvard University, Cambridge, MA 02138} 
 \author{Robert-Jan Slager}
\affiliation{Department of Physics, Harvard University, Cambridge, MA 02138} 
\date{\today}

\begin{abstract}
The possibility of engineering experimentally viable systems that realize 
gauge fluxes within plaquettes of hopping have been subject of search for decades due to vast amounts of theoretical study. This is of particular interest for topological band insulators, where it is known that such fluxes bind protected mid-gap states. These modes can hybridize in extended flux lattices, giving rise to semi-metallic bands that are highly tunable. We demonstrate that within artificial materials, local $\pi$-fluxes can be naturally realized. Consequently, we  provide concrete set-ups to access this physics and analyze similar self-organized band structures and physical properties. Our work therefore does not only pinpoint simple systems exhibiting flat bands, but also opens up a route to study flux lattice models and associated effective theories in a novel scene.
\end{abstract}

\maketitle
    
\paragraph{Introduction-}
Gauge fluxes have been associated with many exotic physical properties studied over several decades, touching upon topics ranging as wide as from e.g visons in lattice gauge theories \cite{RevModPhys.51.659} to quasi-particle descriptions in quantum Hall systems \cite{laughlin1983anomalous}. Similarly, topological band insulators \cite{RevModPhys.82.3045,RevModPhys.83.1057}, arising by virtue of an interplay between symmetry and topology \cite{Clas1a, Clas1b, Clas1c, Clas2, Clas3, 2018wilson, Clas4, PhysRevB.98.024310, Clas5}, or symmetry protected states in general \cite{gu2009tensor, chen2013symmetry,pollmann2012symmetry}, can be unambiguously distinguished from trivial states by monodromy defects \cite{ran2008spin,qi2008spin,PhysRevLett.108.106403,Mes2013,RevModPhys.89.041004}, i.e effective gauge fluxes. Such defects bind modes that relate to the relevant topological invariants.

Despite the richness of such theoretical insights, realizing gauge fluxes in accessible set-ups is still a standing question. The difficulty here mainly lies in the condition that the mentioned model settings require fluxes that are localized to a (or a few) plaquette of hopping sites. Although intriguing proposals utilizing the controllability of cold atom systems \cite{RevModPhys.83.1523} or vortex lattices in type-II superconductors \cite{weeks2007anyons} maybe become more within reach over time, such experiments will unavoidably be rather involved. This in some sense also applies to routes that use lattice defects to act as effective fluxes \cite{ran2009one,imura2011weak,PhysRevLett.108.106403, PhysRevB.90.241403, PhysRevB.93.245406}. While recent experimental results in fact seem promising \cite{bisbdis}, controlling and manipulating defects will require additional progress. 

In this work we analyze physical consequences of localized fluxes within the context of readily controllable artificial materials.  We retrieve $\pi$-flux bound midgap 'zero' modes in topological regimes, as in the electronic counterparts. Apart form providing concrete setups, we show that in extended flux lattices these modes hybridize into midgap band structures that have gapless touching points at specific points in the Brillouin zone. This leads to new transport features of the highly tunable bands that can readily be flattened by increasing the spacing of the flux lattice. Given the implementability of the outlined structures this opens the door to study gauge flux models in a controllable set-ups 
and harvest the associated rich physics. 

\paragraph{Model-}
\begin{figure}[h]
    \centering
    \includegraphics{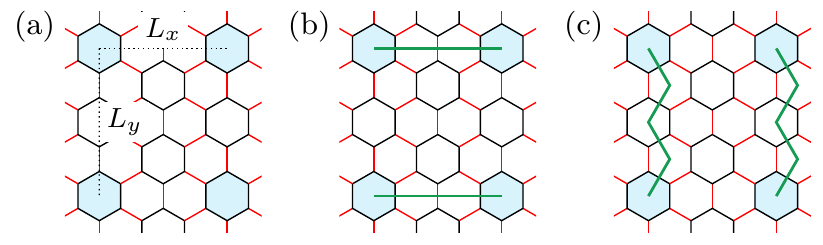}
    \caption{(a) Schematic picture for the modulated honeycomb lattice model with $\pi$-flux. The hopping amplitude is different on the black and the red bonds. The shaded hexagons are threaded by $\pi$-flux. (b,c) Corresponding $\pi$-flux strings. The strings in (b) and (c) are gauge related.}
    \label{fig:schematic}
\end{figure}
To make our arguments concrete, we analyze band dispersions of a modulated honeycomb lattice tight binding model with pi-flux lattice on top of it. We consider the simple Hamiltonian
\begin{equation}
H=\sum_{\langle ij\rangle}t_{ij}c^{\dagger}_{i}c_{j},
\end{equation}
where $\langle ij\rangle$ refers to the nearest neighbours and $c^{\dagger}_i$ is the creation operator at site $i$. Furthermore, the hopping parameters $t_{ij}$ are set in terms of modulations of the type depicted in Fig.~\ref{fig:schematic}, in which we assign intra-hexagon hopping $t_0=t+\delta$ for black bonds and inter-hexagon hopping $t_1=t-\delta$ for red bonds. This modulation gaps out Dirac cones of the original honeycomb lattice model through the coupling between the Dirac cones at the K and K' points in the Brillouin zone \cite{PhysRevLett.114.223901,Wu:2016aa,Kariyado:2017aa}. Importantly, the states near zero energy are described by a massive Dirac equation where the sign of the mass term can be flipped by changing the ratio between $|t_0|$ and $|t_1|$, or the sign of $\delta$. The feature that the sign flip of the mass term can be induced by a little modulation in the hopping amplitude makes this model an ideal playground to simulate a quantum spin Hall state in artificial systems. Indeed, several of the experimental demonstrations of helical edge states in the literature in fact implement this model \cite{Yves:2017aa,Barik666,Yu:2018aa,Li:2018aa,Noh:2018aa,PhysRevLett.120.217401,Zhirihin_2018,Chaunsali_2018}. In the following, we set $t=-1$, and $|\delta| = 0.1$. Note that $\delta>0$ corresponds to $|t_0|<|t_1|$ while $\delta<0$ corresponds to $|t_0|>|t_1|$. Throughout the paper, we use $a_0$ as the nearest neighbor distance between the sites in the honeycomb lattice.

To implement a $\pi$-flux lattice, it is convenient to think of the conjugate expression by strings. Namely, we pick up two hexagons and connect them by a string, and every time the string hits the bond, we flip the sign of the hopping on that bond. Then we are left with two $\pi$-flux threaded hexagons at the ends of the string. There is arbitrariness in the path of the strings, i.e., if two strings form a loop when they are combined, the two ways are related by a gauge transformation and physically equivalent. For instance, Figs.~\ref{fig:schematic}(b) and \ref{fig:schematic}(c) are physically the same. As in the case of the uniform magnetic field, in order to make the Hamiltonian strictly periodic (for calculating band structures), the supercell of the $\pi$-flux lattice has to contain an integer multiple of full flux ($2\pi$), indicating that the supercell should contain two $\pi$-flux threaded hexagons. In the following, when we see the band structure as a function of $k_x$, we implicitly chose the string alignment as in Fig.~\ref{fig:schematic}(c) to minimize the period in $x$-direction, while when we see the band structure as a function of $k_y$, the alignment in Fig.~\ref{fig:schematic} is chosen to minimize the period in $y$-direction.

\begin{figure}
    \centering
    \includegraphics{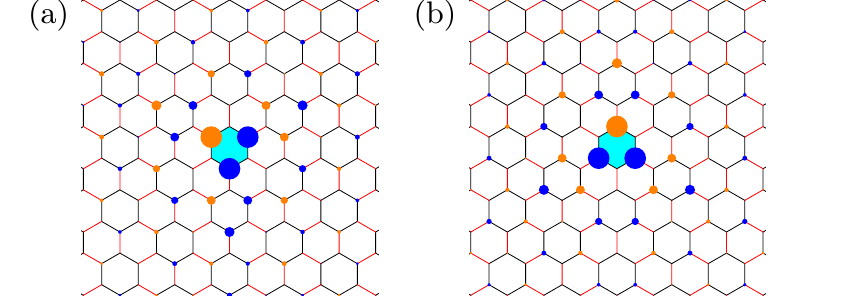}
    \caption{Wave functions for the two midgap states bound to an isolated $\pi$-flux marked as a cyan hexagon. The size of each of the disks represents the amplitude of the wave function at the corresponding site, while the color (orange or blue) is for the sign of the wave function. $L_x=36\sqrt{3}a_0$ and $L_y=72a_0$ are adapted and the gauge in Fig.~\ref{fig:schematic}(c) is used.}
    \label{fig:wf}
\end{figure}
Because of the hopping modulation, hexagons in the current model can be classified into two kinds, one consists of the black bonds exclusively, while the other consists of the mixture of the black and the red bonds. We first consider the case that the $\pi$ fluxes are placed on the hexagon of the first kind, as in Fig.~\ref{fig:schematic}, and check the existence of the midgap states by looking at the isolated $\pi$-flux. In practice, we numerically obtain the energy spectrum and the wave functions with sufficiently large $L_x$ and $L_y$. A clear sign of the midgap state is found for $|t_0|>|t_1|$. There appear two bound states per a $\pi$-flux, whose wave functions are shown in  Fig.~\ref{fig:wf}.

\begin{figure}
    \centering
    \includegraphics{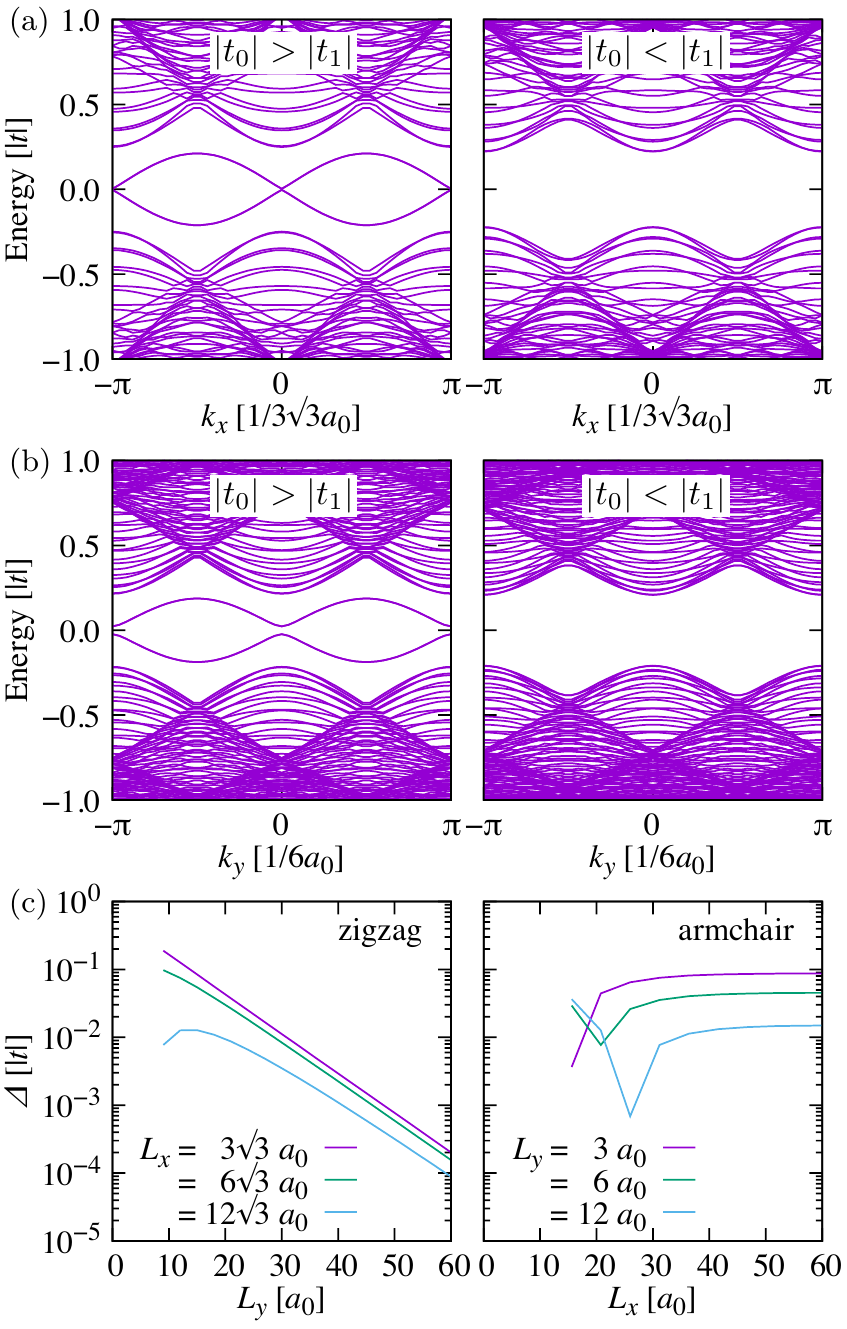}
    \caption{(a) Band structure for $L_x=3\sqrt{3}a_0$ and $L_y=36a_0$. With this setup, the $\pi$-flux threaded hexagons essentially align in the zigzag direction, and the focus is on the band dispersion along that direction. (b) Band structure for $L_x=12\sqrt{3}a_0$ and $L_y=6a_0$. The focus is on the band dispersion along the armchair direction. (c) Energy gap across the zero energy at $k_{x,y}=0$. Left: $L_y$ dependence for several choices of $L_x$, proving the gapless nature. Right: $L_x$ dependence for several choices of $L_y$, detecting finite mini gap not from the size effect.}
    \label{fig:bands}
\end{figure}
For the case with $L_x\ll L_y$, the system is essentially regarded as a one-dimensional $\pi$-flux lattice ($\pi$-flux chain) along the zigzag direction, while for the case  with $L_x\gg L_y$, the system is regarded as a $\pi$-flux chain along the armchair direction. In such cases, we can think of band structures generated by hybridizations between the bound states of the neighboring $\pi$-fluxes.
The results for $L_x\ll L_y$, i.e., the band structures of the $\pi$-flux chain along the zigzag direction are shown in Fig.~\ref{fig:bands}. We clearly see an in-gap state with 1D Dirac structure at the zone center and the corner for $|t_0|>|t_1|$, while no in-gap state for $|t_0|<|t_1|$ is retrieved. Similar band structures are obtained also for $L_x\gg L_y$, i.e., for the $\pi$-flux chain along the armchair direction. However, in the armchair case, we numerically obtain a mini gap within the in-gap band at the zone center and boundary (k=$\pi$). Figure~\ref{fig:bands}(c) shows the gap $\Delta$ across zero energy at the zone center. For the chain along the zigzag direction, $\Delta$ exponentially decays as function of the interchain distance $L_y$, which signals gapless nature of the single chain. On the other hand, for the chain along the armchair direction, $\Delta$ saturates as a function of the interchain distance $L_x$, conveying that the mini gap is not coming from size effects.

\begin{figure}
    \centering
    \includegraphics{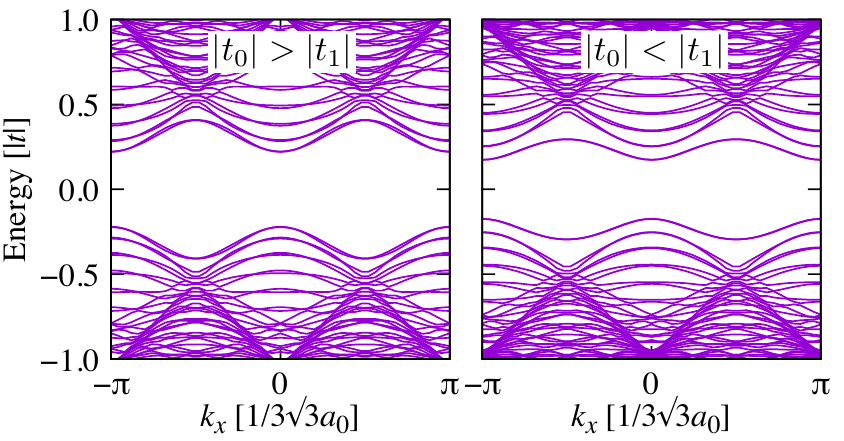}
    \caption{Band structure for the case that the $\pi$-fluxes are placed on the hexagons consisting of the black and the red bonds. $L_x=3\sqrt{3}a_0$ and $L_y=36a_0$ are adapted.}
    \label{fig:2nd_kind}
\end{figure}
The $\pi$-flux may also be put on the hexagons of the second kind, for instance by shifting the strings in Figs.~\ref{fig:schematic}(b) or \ref{fig:schematic}(c) by the size of a single hexagon. The band structures for the $\pi$-flux chain threading the hexagons of the second kind are shown in Fig.~\ref{fig:2nd_kind}. We see that some states pop up (and down) from the bulk continuum to the gap for $|t_0|<|t_1|$, but there is no clear sign of zero gap or near zero gap states. A consideration on the extreme limit of $t_0=0$ or $t_1=0$ gives some insights for the difference between threading the hexagons of the first and the second kind, and also the difference between $|t_0|>|t_1|$ and $|t_0|<|t_1|$. If we have $t_1=0$, the system is a set of isolated hexagons of the first kind. In such a case, it is easy to convince ourselves that the $\pi$-flux threading of those hexagons gives rise to zero energy states, and nothing happens if the $\pi$-fluxes are misplaced. On the other hand, if we have $t_0$, the system is a set of dimers with no loops, making $\pi$-fluxes ineffective. Whether or not the zero energy state in the extreme limit survives in the interested parameter region is a nontrivial issue, and topology and symmetry come into play to assist the zero energy state to survive, as we will see in the following.

\paragraph{Stability-}
The stability of the $\pi$-flux modes finds its origin in the existence of edge states and thus hangs directly together with the topological phase \cite{ran2008spin,qi2008spin,
PhysRevB.92.085126, PhysRevB.97.115143, ran2009one, PhysRevB.93.245406, imura2011weak}. In fact they provide for a direct bulk probe of the non-trivial symmetry protected topology. As outlined in \cite{ran2009one}, this can readily seen using a Volterra argument. Specifically, one can consider the effect of a $\pi$-flux by hypothetically cutting the system and then gluing back the two sides after the hopping elements over the segment that connects the boundary and defect core are multiplied with a $\pi$-phase, $e^{i\pi}$. Considering the case when the cutting the system results in edge states, gluing the sides together conventionally results in a mass term gapping the spectrum back to the original bulk situation. However, as the phases induce a minus sign for the hoppings on one side of the defect core this mass term now switches sign. As result, we obtain an effective Dirac system with changing mass term,
\begin{equation}
	H_0 = v  k \sigma_3 \mu_3 + m(x) \mu_1, \label{eq:glue}
\end{equation}
where $v$ specifies the velocity, $\mu$ refers to the two edges of the Volterra cut and $\sigma$ parametrizes the pseudo spin. Such Dirac theories with changing mass terms are well known to host localized soliton states at the mass domain wall, being the $\pi$-flux mode.  

Our remaining task is to show the stability of the gapless edge mode supporting the description by Eq.~\eqref{eq:glue}. Here we give two complementary explanations, one using a pseudo time reversal symmetry (TRS) and another using a mirror winding number. 
As for the former one, the system at hand is characterized by a $\mathbf{Z}_2$ classification, which is associated with a pseudo time reversal symmetry (TRS), see also Supplemental Material. In particular, time reversal, represented by complex conjugation ${\cal K}$, can be lifted to an anti-unitary operation under $C_n$ point group symmetries for $n=4,6$ \cite{Clas1c}. The bands with complex values under the rotation operator associated with the $E$ representation, as for e.g. as $|p_x\rangle$ and $|p_y\rangle$ states, come in pairs and induce a {\it two dimensional  real} representation due to the combination of both symmetries leading to degeneracies at the $\Gamma$ and $K$ points in the Brillouin zone. These doublets can then give rise to a $\mathbf{Z}_2$ non-trivial phase by 'switching', rather similar to spinful TRS systems. This mechanism is also at work in the present context, where doublet at the $\Gamma$ point, arising by virtue of the six-fold symmetry give rise to a non-trivial $\mathbf{Z}_2$ invariant. This argument only relies on doublets ($E$ representation) and has way broader applicability than our specific model, however we should note that this TRS comes from the crystal symmetry and only emerges in the effective model. In fact the edge states can be gapped out by continuing away from the $\Gamma$-point in the Brillouin zone, showing that the robustness of this symmetry is indeed not generally to be taken for granted. 

Alternative, more rigorous,  support for the gapless edge modes at zigzag edges comes from the combination of the chiral (sublattice) symmetry and the mirror symmetry, whose reflection plane is perpendicular to the zigzag edge \cite{Kariyado:2017aa,PhysRevB.98.195416}. The gaplessness comes from the degeneracy of the two zero modes having opposite chirality (sublattice polarization) and the opposite parity with respect to the reflection, guaranteed by nontrivial mirror winding numbers. In contrast to the pseudo TRS based argument, the mirror winding number is completely robust, although the chiral symmetry is generically a strong constraint.

Finally, we turn to the gaplessness (mini gap) of the midgap bands for the zigzag (armchair) $\pi$-flux chain, by focusing on the subspace spanned by the $\pi$-flux states. The above arguments suggests that the $\pi$-flux state inherits the properties of the edge states, and the two bound states at a single $\pi$-flux have opposite chiralities (see Fig.~\ref{fig:wf}). Then, within the effective model, we can take a basis such that the chiral operator $\gamma$ becomes $\gamma=\sigma_3$, which gives us a chiral symmetric effective Hamiltonian
\begin{equation}
 H(\tilde{k}) = h_1(\tilde{k})\sigma_1+h_2(\tilde{k})\sigma_2,
\end{equation}
where $\tilde{k}=k_x$ ($\tilde{k}=k_y$) for the zigzag (armchair) $\pi$-flux chain (see Fig.~\ref{fig:schematic}), referring to $x,y$ direction whereas the value is set by the spacing of fluxes. For the zigzag $\pi$-flux chain, we also have the reflection symmetry $\sigma_3 \hat{I}$, where $\hat{I}$ brings $\tilde{k}$ to $-\tilde{k}$, which commutes with $\gamma$ at $\tilde{k}=0$, see Supplemental Material. (Note that under the influence of the $\pi$-fluxes, the reflection symmetry actually means the spacial reflection associated with a proper gauge transformation.) This gives $h_{1,2}(-\tilde{k})=-h_{1,2}(\tilde{k})$, and retaining the nearest neighbor hoppings, the band structure will result in $E(\tilde{k})\sim\pm|\sin \tilde{k}|$. On the other hand, for the armchair $\pi$-flux chain, the reflection symmetry becomes $\sigma_1 I$, which anticommute with $\gamma$ (see Supplemental Material), giving us $h_1(-\tilde{k})=h_1(\tilde{k})$ and $h_2(-\tilde{k})=-h_2(\tilde{k})$. Taking only the leading contribution to $h_{1,2}(\tilde{k})$, 
the dispersion relation becomes $E(\tilde{k})=\pm\sqrt{m^2+\tilde{t}^2\sin\tilde{k}^2}$. These arguments perfectly agree with the numerical results. As a side note, we mention that these considerations can also be phrased using the domain wall argument. That is, one can readily formulate all, symmetry consistent, couplings  between the Volterra cuts and then project down to the subspace of the soliton solutions. These argument then lead to the same effective Hamiltonian. Generalizing this description to account for the different symmetries involved for the zigzag and armchair edge then similar undertsnading of the diffrence between these cases, while this set-up also allows for the addition of other (TRS) symmetry breaking terms and study those effects \cite{PhysRevB.93.245406}.

\begin{figure}
    \centering
    \includegraphics{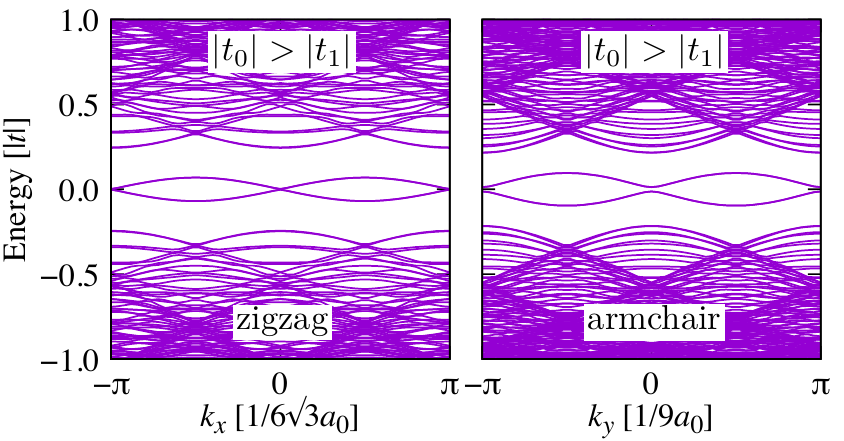}
    \caption{Flatter in-gap bands for larger $\pi$-flux spacing. Left (Right) is for $L_x=6\sqrt{3}a_0$ ($12\sqrt{3}a_0$) and $L_y=36a_0$ ($9a_0$), where the $\pi$-flux threaded hexagons align in the zigzag (armchair) direction.}
    \label{fig:flatter}
\end{figure}

\paragraph{Perspective-}
Our demonstration suggests $\pi$-flux as a nice building block for designing various kinds of states. For instance, since the $\pi$-flux state is localized,
 simply making the distance between the threaded hexagons away results in flattened bands. Figure \ref{fig:flatter} shows such examples, i.e., we observe smaller total band width of the in-gap states in the case of larger distance between the threaded hexagons. For the chains along the armchair direction, the mini gap is also reduced, which has also been detected in  Fig.~\ref{fig:bands}(c). The advantage of using $\pi$-fluxes as building blocks is that the existence of zero energy states are guaranteed by {\it topological protection}. In addition, in the same modulated honeycomb lattice model, edge states or interface states at the boundary between two regions with $|t_0|>|t_1|$ and $|t_0|<|t_1|$ have been investigated. Hence, it is an interesting future problem to look for new states or phenomena caused by interplay between edge/interface states and $\pi$-flux states.

\paragraph{Experimental implementation-} 
Obviously, the key ingredient to realize $\pi$-flux is feasibility of the sign flip of the hopping. In some artificial systems, schemes to change the sign of the coupling have already been proposed. For instance, Ref.~\cite{doi:10.1063/1.4942357} proposes a mechanical system made of disks and springs, and there ``normal'' and ``reversed'' springs leading to opposite signs of couplings are introduced. Sometimes, spring-based discrete systems are effectively or essentially realized in continuous elastic media with springs being substituted by beams. Mapping a discrete model to a continuous model is a nontrivial task, but once the mapping is established, current 3D printing techniques will facilitate us to fabricate a designed elastic media \cite{Wang_2018}. Another possible and promising scheme is to use LC circuits. Ref.~\cite{PhysRevB.99.020302} proposes a clever wiring between elements to integrate 0-phase  and $\pi$-phase hoppings into a single system, and furthermore, the authors of Ref.~\cite{PhysRevB.99.020302} have already implemented a real circuit with both 0-phase and $\pi$-phase hoppings to show Berry curvatures and Fermi arcs.

Even in a system where tuning of hoppings itself is difficult, if we have control over site potentials instead, it is in principle possible to build a system with effective $\pi$-flux. A rough sketch of the scheme is the following. Let us focus on two sites across which we want to induce hopping. Then, instead of directly connecting them, we put an extra intermediate site connected to the both of the focused sites by hopping $t$. If we set the difference of the site potentials for the intermediate and the focused sites $\Delta E$ large enough, the second order perturbation leads to an effective hopping between the focused site $t^2/\Delta E$, which means that the sign of the effective hopping can be chosen by changing the sign of $\Delta E$.

\paragraph{Conclusion and Discussion-}
We have shown that $\pi$-flux modes, characterizing topological band insulators, can be naturally realized in artificial materials. Given the {\it direct} implementabilty of artificial materials,  this creates a promising scene to study $\pi$-flux modes as well as the highly tuneable bands that arise from their hybridization. More generically, these modes can be used as building blocks, opening up the door to an experimentally viable setting in which numerous models incorporating fluxes (vortex lattices,  lattice gauge theories etc.) and the associated effective (field) theories can be brought to life. 

We note that in contrast to electron systems, such bosonic set-ups do not have a natural filling, constraining the possibilities of anomalies that also arise due to an interplay with the edge states of the parent model \cite{PhysRevB.93.245406}. However, the midgap modes still can be viewed as extra conducting channels, which will have graphene-like signatures (or in fact half graphene like signatures due to the absence of a valley degree of freedom). Indeed, as alluded to above, an interesting direction will be to study the interaction with edge states, considering extended flux lattices connecting edges. This similarly applies to studying junctions of flux lattices in the bulk where the modes can form new transporting architectures.  Finally, we also intend to study dense flux lattices in which the bound modes can hybridze into novel band structures that can have (higher) Chern numbers depending on the coordination of this flux lattice. Motivated by these results and novel possible routes, we hope that our work will inspire a range of future pursuits.

\paragraph{Acknowledgements} R.-J.S appreciatively acknowledges funding via Ashvin Vishwanath from the Center for the Advancement of Topological Materials initiative, an Energy Frontier Research Center funded by the U.S. Department of Energy, Office of Science. This work was partially supported by JSPS KAKENHI Grant Numbers JP17K14358 and JP18H01162 (T.K.).                      

\appendix
\section{Appendix}
\subsection{Symmetry}
\begin{figure}[btp]
 \begin{center}
  \includegraphics{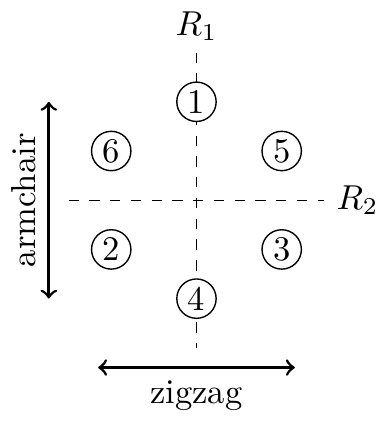}
  \caption{Labeling of the sublattices in a unit cell. Reflection planes $R_1$ and $R_2$ are also shown.}\label{fig:sublattices}
 \end{center}
\end{figure}

In this Supplemental Material, the symmetries for the model are detailed. 
Using the unit vectors $\bm{a}_1={}^t(3\sqrt{3}a_0/2,3a_0/2)$ and $\bm{a}_2={}^t(-3\sqrt{3}a_0/2,3a_0/2)$, the Hamiltonian of our model is written as
\begin{equation}
    H(k_x,k_y)=
    t_0
  \begin{pmatrix}
   0 & D \\
   D^\dagger & 0 
  \end{pmatrix},\quad
   D=
  \begin{pmatrix}
   \alpha e_1^*e_2^* & 1 & 1 \\
   1 & \alpha e_1 & 1\\
   1 & 1 & \alpha e_2
    \end{pmatrix},
\end{equation}
where the sites in a unit cell is labeled as Fig.~\ref{fig:sublattices}, $e_l=e^{i\bm{k}\cdot\bm{a}_l}$ and $\alpha=t_1/t_0$. This Hamiltonian has (i) the chiral symmetry
\begin{equation}
 H(k_x,k_y)=-\gamma H(k_x,k_y) \gamma,\quad \gamma^2=1.
\end{equation}
with
\begin{equation}
 \gamma = 
  \begin{pmatrix}
   1 & 0 & 0 & 0 & 0 & 0 \\
   0 & 1 & 0 & 0 & 0 & 0 \\
   0 & 0 & 1 & 0 & 0 & 0 \\
   0 & 0 & 0 &-1 & 0 & 0 \\
   0 & 0 & 0 & 0 &-1 & 0 \\
   0 & 0 & 0 & 0 & 0 &-1
  \end{pmatrix},
\end{equation}
(ii) the mirror symmetry with the reflection plane perpendicular to the zigzag edge
\begin{equation}
 H(-k_x,k_y) = R_1 H(k_x,k_y) R_1,
\end{equation}
and (iii) the mirror symmetry with the reflection plane perpendicular to the armchair edge
\begin{equation}
 H(k_x,-k_y) = R_2 H(k_x,k_y) R_2
\end{equation}
where 
\begin{equation}
 R_1 = 
  \begin{pmatrix}
   1 & 0 & 0 & 0 & 0 & 0 \\
   0 & 0 & 1 & 0 & 0 & 0 \\
   0 & 1 & 0 & 0 & 0 & 0 \\
   0 & 0 & 0 & 1 & 0 & 0 \\
   0 & 0 & 0 & 0 & 0 & 1 \\
   0 & 0 & 0 & 0 & 1 & 0
  \end{pmatrix},\quad
  R_2 = 
  \begin{pmatrix}
   0 & 0 & 0 & 1 & 0 & 0 \\
   0 & 0 & 0 & 0 & 0 & 1 \\
   0 & 0 & 0 & 0 & 1 & 0 \\
   1 & 0 & 0 & 0 & 0 & 0 \\
   0 & 0 & 1 & 0 & 0 & 0 \\
   0 & 1 & 0 & 0 & 0 & 0
  \end{pmatrix}.
\end{equation}
Note that $\gamma$ and $R_1$ are compatible since $[\gamma,R_1]=0$, while $\gamma$ and $R_2$ are incompatible since $\{\gamma,R_2\}=0$.
To derive an effective model near the $\Gamma$-point and the zero energy, we focus on the subspace spanned by a new basis set $\{|u_-\rangle,|u_+\rangle,|l_-\rangle,|l_+\rangle\}$ defined as
\begin{equation}
    |u_\pm\rangle=
        \begin{pmatrix}
        |\pm\rangle\\
        0
        \end{pmatrix},\quad
    |l_\pm\rangle=
        \begin{pmatrix}
        0\\
        \pm|\pm\rangle
        \end{pmatrix}
\end{equation}
and
\begin{equation}
     |\pm\rangle = \frac{1}{\sqrt{3}}
  \begin{pmatrix}
   1\\
   \omega_\pm\\
   \omega_\mp
    \end{pmatrix},\quad
    \omega_\pm = -\frac{1}{2}\pm\frac{\sqrt{3}}{2}i,
\end{equation}
which gives us
\begin{equation}
  H(k_x,k_y) = t_0 
   \begin{pmatrix}
    0 & \tilde{D} \\
    \tilde{D}^\dagger &0 \\
   \end{pmatrix}
\end{equation}
with 
\begin{equation}
 \tilde{D} = 
   \begin{pmatrix}
   -\langle -|D|-\rangle & \langle -|D|+\rangle \\
   -\langle +|D|-\rangle & \langle +|D|+\rangle
  \end{pmatrix}.
\end{equation}
We have, up to the second order in $|k|$,
\begin{equation}
 \langle +|D|+\rangle
 = \langle -|D|-\rangle
 = -(1-\alpha)-\alpha \frac{9a_0^2}{4}(k_x^2+k_y^2),
\end{equation}
and up to the first order in $|k|$, 
\begin{equation}
 \langle  -|D|+\rangle = \alpha\frac{3a_0}{2}k_-,\quad
 \langle  +|D|-\rangle = -\alpha\frac{3a_0}{2}k_+,
\end{equation}
leading to
\begin{multline}
 H(k_x,k_y) = \Bigl((t_0-t_1)+t_1\frac{9a_0^2}{4}(k_x^2+k_y^2)\Bigr)\sigma_3\otimes\mu_1\\
 + t_1 \frac{3a_0}{2} (\bm{k}\cdot \bm{\sigma})\otimes\mu_1.
\end{multline}
If we further go to the basis set
\begin{equation}
 \frac{|u_-\rangle+|l_-\rangle}{\sqrt{2}},\,
 \frac{|u_+\rangle+|l_+\rangle}{\sqrt{2}},\,
  \frac{|u_+\rangle-|l_+\rangle}{\sqrt{2}},\,
 \frac{|u_-\rangle-|l_-\rangle}{\sqrt{2}},\label{eq:basis1}
\end{equation}
the Hamiltonian becomes
\begin{equation}
 H(k_x,k_y) = 
 \begin{pmatrix}
    H_+(k_x,k_y) & 0 \\
    0 & H_-(k_x,k_y)
 \end{pmatrix},
\end{equation}
with
\begin{multline}
H_\pm(k_x,k_y)=\Bigl((t_0-t_1)+t_1\frac{9a_0^2}{4}(k_x^2+k_y^2)\Bigr)\sigma_3\\
 + t_1 \frac{3a_0}{2} (k_x\sigma_1\pm k_y\sigma_2).
\end{multline}
Each of $H_+(k_x,k_y)$ and $H_-(k_x,k_y)$ has the (pseudo) time reversal symmetry $i\sigma_2 \mathcal{K}$ with $\mathcal{K}$ being complex conjugation.

Using, 
\begin{align}
 |p_x\rangle&=\frac{1}{2}
  \begin{pmatrix}
   0\\
   1\\
  -1\\
   0\\
   1\\
  -1
  \end{pmatrix},\quad
 |p_y\rangle&=\frac{1}{2\sqrt{3}}
  \begin{pmatrix}
   2\\
  -1\\
  -1\\
  -2\\
   1\\
   1
  \end{pmatrix},\\
 |d_{x^2-y^2}\rangle&=\frac{1}{2\sqrt{3}}
  \begin{pmatrix}
  -2\\
   1\\
   1\\
  -2\\
   1\\
   1
  \end{pmatrix},\quad
 |d_{xy}\rangle&=\frac{1}{2}
  \begin{pmatrix}
   0\\
  -1\\
   1\\
   0\\
   1\\
  -1
  \end{pmatrix},
\end{align}
the basis set Eq.~\eqref{eq:basis1} is rewritten as
\begin{gather}
 i\frac{|p_x\rangle-i|p_y\rangle}{\sqrt{2}},\quad
 -\frac{|d_{x^2-y^2}\rangle-i|d_{xy}\rangle}{\sqrt{2}},\\
 -i\frac{|p_x\rangle+i|p_y\rangle}{\sqrt{2}},\quad
 -\frac{|d_{x^2-y^2}\rangle+i|d_{xy}\rangle}{\sqrt{2}},
\end{gather}
namely, 
\begin{equation*}
 i|p_-\rangle,\quad -|d_-\rangle,\quad -i|p_+\rangle,\quad -|d_+\rangle.
\end{equation*}

By definition, we have the following relations,
\begin{gather}
 \gamma |p_x\rangle = -|d_{xy}\rangle,\quad
  \gamma |p_y\rangle = -|d_{x^2-y^2}\rangle,\\
  \gamma |d_{x^2-y^2}\rangle = -|p_y\rangle,\quad
  \gamma |d_{xy}\rangle = -|p_x\rangle,\\
 R_1 |p_x\rangle = -|p_x\rangle,\quad
 R_1 |p_y\rangle =  |p_y\rangle,\\
 R_1 |d_{x^2-y^2}\rangle = |d_{x^2-y^2}\rangle,\quad
 R_1 |d_{xy}\rangle = -|d_{xy}\rangle,\\
 R_2 |p_x\rangle = |p_x\rangle,\quad
 R_2 |p_y\rangle = -|p_y\rangle,\\
 R_2 |d_{x^2-y^2}\rangle =|d_{x^2-y^2}\rangle,\quad
 R_2 |d_{xy}\rangle = -|d_{xy}\rangle,
\end{gather}
which lead to
\begin{equation}
 \gamma = \sigma_1\otimes\mu_1, \quad 
 R_1 = \hat{1}\otimes \mu_1,\quad
 R_2 = -\sigma_z\otimes \mu_1.
\end{equation}

\subsection{Symmetry with $\pi$-flux}
\begin{figure}
    \centering
    \includegraphics{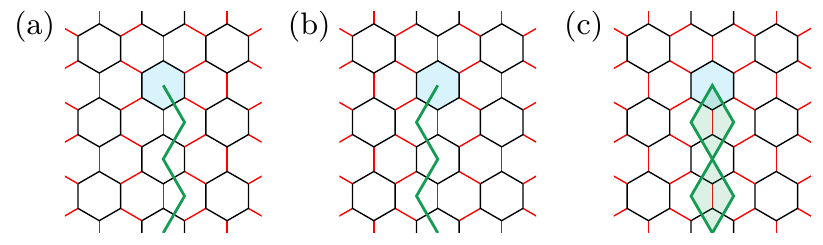}
    \caption{(a) Original configuration. (b) Only the spatial reflection is applied. (c) Schematic description of the gauge transformation. The sign of the basis function is flipped in the shaded region.}
    \label{fig:gauge_trans}
\end{figure}

With $\pi$-fluxes or $\pi$-phase strings, special care is required regarding the reflection symmetry. In short, we have to augment the spatial reflection by a gauge transformation. Let us consider to apply the reflection along the vertical line on the configuration in Fig.~\ref{fig:gauge_trans}. The spatial reflection alone brings Fig.~\ref{fig:gauge_trans}(a) to Fig.~\ref{fig:gauge_trans}(b), which is not identical to the original configuration. If we further apply the gauge transformation such that the sign of the wave function within the shaded region in Fig.~\ref{fig:gauge_trans}(c) is flipped, Fig.~\ref{fig:gauge_trans}(b) goes back to Fig.~\ref{fig:gauge_trans}(a). Note that this in fact amounts to a large gauge transformation of $2\pi$, as conveyed by combining the paths in the right column, and that this indeed shows the reflection symmetry of the states in Fig.~2 in the main text. Equipped with this combined operation of the spatial reflection and the gauge transformation, it is straightforward to convince ourselves that the wave function in Fig.~2(a) in the main body is odd against this symmetry operation, while the one in Fig.~2(b) is even against this symmetry operation. This legitimate us to use $\sigma_z \hat{I}$ as a reflection symmetry within the subspace spanned by the $\pi$-flux states. 

\end{document}